\documentstyle[11pt]{article}
\input epsf
\begin{document}
\baselineskip 18pt
\rightline{CU-TP-702}
\vskip 1cm
\centerline{\Large\bf QUENCHED HAWKING RADIATION \&}
\centerline{\Large\bf THE BLACK HOLE PAIR-CREATION RATE\footnote{to appear in
the proceedings of the Sixth Moscow Quantum Gravity Seminar, June 12-19, 1995}}
\vskip 1cm
\centerline{{\sc Piljin Yi} \footnote{e-mail address:
piljin@cuphyc.phys.columbia.edu}}
\vskip 5mm
\centerline{\it Physics Department, Columbia University}
\centerline{\it 538 West 120th St., New York, New York, 10027, U.S.A}
\vskip 1cm
\begin{quote}
{\small
The main topic of this talk is the Hawking effect when the black holes in
question are undergoing a uniform acceleration. The semiclassical effect of
the acceleration is most striking when the Hawking temperature equals the
acceleration temperature. Within the usual late-time approximation, the
natural black hole vacuum is then equivalent to the asymptotically empty
Minkowskian vacuum, and the accelerated black hole becomes semiclassically
stable against the familiar thermal evaporation. An important application
of this phenomenon is found in the problem of charged black
hole pair-creation.}
\end{quote}

\vskip 1cm

The problem I wish to address in this talk is essentially that of the
Hawking effect in the presence of uniformly accelerated black holes.
Without the acceleration, the natural vacua
of the black holes are well known to be of thermal type with respect to the
asymptotic inertial observers who must then find a blackbody radiation from
the physical black holes \cite{hawking}. When the black
holes are undergoing acceleration, however, two important facts invalidate
this conventional semiclassical picture. First of all, the vacuum would be
again thermal with respect to the {\it asymptotic co-moving} observers who are
following the black hole at a large fixed distance, but they are no longer
{\it inertial} observers.  One must reidentify the asymptotic inertial
observers and determine how they would perceive the black hole vacuum.
Second, even for those {\it asymptotic co-moving} observers, the Hawking
radiation could be modified because of another related, if less publicized,
quantum effect known as Fulling-Davies-Unruh effect or the acceleration heat
bath \cite{bath}\cite{unruh}.

A convenient starting point of this discussion would be this
Fulling-Davies-Unruh effect. The statement is that, if any
quantum particle detection device is undergoing a uniform acceleration, it
will register the presence of a heat bath of particles at a temperature, call
it $T_A$, proportional to the absolute acceleration. This is often expressed by
saying that the the usual Minkowski vacuum feels like a heat bath to any
accelerated observers. However, one should keep in mind that the state is
still the Minkowskian vacuum, so that, for example, the covariant
energy-momentum expectation values are trivial despite the thermal
characteristics.

With this in mind, we may consider the following gedanken experiment
\cite{letter}\cite{giddings}: Let a black hole of the Hawking temperature
$T_{BH}$ be accelerated uniformly such that the acceleration temperature
$T_{A}$ equals $T_{BH}$. What happens to the semiclassical evolution
of the black hole? Does the black hole still Hawking-radiate and
thereby lose its mass continuously? Or will the acceleration heat bath
counteract the Hawking flux and stabilize the system semiclassically?
Going one more step, one may further ask if it is possible for the black hole
to Hawking-radiate toward asymptotic inertial observers while being in an
apparent thermal equilibrium with the acceleration heat bath.

A comparison to a classic puzzle illustrates well how nontrivial this problem
is in principle. Suppose we consider a charged particle under a uniform
acceleration {$\bf A$}. According to the classical eletrodynamics, then, the
charge emits the classical Bremmstrahlung, the power of which is given
by the following formula,
\begin{equation}
\frac{dE}{dt}=\frac{2e^2}{3}\,{\bf A}^2.
\end{equation}
But the exactly same field
configuration, if seen by {\it co-moving} accelerated observers, appears
as the static Coulomb field of $1/r^2$ tail without a hint of radiative
behavior. Furthermore, the uniformly accelerated charge does not experience
the familiar radiation damping force that normally converts  the kinetic
energy of the charge to the radiation energy of the Bremmstrahlung:
\begin{equation}
{\bf F}_{\rm damping}=\frac{2e^2}{3}\,\dot{\bf A}\rightarrow 0 ,
\end{equation}
and this leads to an apparent discrepancy with the energy conservation. This
famous puzzle had been an outstanding issue for several decades in this
century, until the resolution was brought under the light in a beautiful work
by Boulware \cite{boulware}, where he explained how all these observations
are actually consistent with each other as well as with the energy-momentum
conservation.

A quantum description \cite{quantum}
of this classical system makes the comparison even more
compelling. Quantum mechanically, the co-moving observers simply find a
detailed balance between the charge and its surroundings (that include the
long range Coulomb field as well as the heat bath of thermal photons) just as
their counterpart in the black hole problem may find a detailed balance
between the two thermal behaviors owing to the fine-tuned
temperatures $T_{BH}=T_{A}$.

As it turned out \cite{letter} and as will be discussed in detail below,
however, the quantum problem associated with
the accelerated black holes above is conceptually
much simpler than this classical counterpart. In a nutshell , the Hawking
radiation is indeed counteracted by the acceleration heat bath which then
prevents thermal evaporation of the black hole. Furthermore, the asymptotic
inertial observers agree with the co-moving observers that the black hole
does not evaporate at all even though the acceleration heat bath is completely
fictitious to the former \cite{letter}. The main purpose of this talk is to
derive and illustrate this hitherto unknown quantum effect and also
consider its implication.

The property of vacua around such
uniformly accelerated black holes is also of some importance
in another important context. As studied in recent years, there
exist  exact solutions to the Euclidean Einstein-Maxwell theory that describe
a uniformly accelerated black hole and that play the
role of the instanton  for the pair-creation of non-extremal
Reissner-Nordstrom black holes via quantum tunneling \cite{garfinkle}.
Such instanton solutions, if continued to the Lorentzian signature,
correspond to the so-called Ernst spacetime where the oppositely charged
Reissner-Nordstrom black holes are uniformly accelerated away from
each other by an external electromagnetic field along a symmetry axis.
Furthermore, provided that $T_{BH}\neq 0$, the Euclidean solution exists
only when the accelerated temperature
matches the Hawking temperature exactly. The effect of quantum fluctuation to
the pair-creation process cannot be understood without the knowledge of the
natural vacua for such accelerated geometry \cite{piljin}.

The rest of the talk is organized as follows. I will start with this Ernst
metric \cite{ernst} as an idealized model for the  gedanken experiment.
After identifying key features of this spacetime, the causal structure
and the asymptotic inertial time coordinate are discussed in comparison
with those of a freely falling black hole. A brief review of the Hawking
effect and the related Bogolubov transformation is outlined, which are then
applied to the current problem. The semiclassical physics of the black hole
horizon and the acceleration horizon are shown to be surprisingly similar,
and this will lead to the main conclusion of this talk. With this new insight,
I will turn to the final topic of the talk: the  one-loop WKB estimate of
black hole pair-creation rate.

\vskip  5mm

Let me first write down the Ernst metric:
\begin{eqnarray}
& &g=\frac{\Lambda^2}{(1+rAx)^2}\biggl\{-F(r)\,ds^2+F(r)^{-1}dr^2+r^2\,
G(x)^{-1}dx^2+r^2\,G(x)\,\Lambda^{-4}d\phi^2\biggr\}, \nonumber\\
& &F(r)\equiv -A^2r^2\,G(-1/Ar)=(1-\frac{r_-}{r})(1-\frac{r_+}{r}-A^2r^2),
\nonumber\\
& &\Lambda\equiv \biggl\{1+\frac{Bx}{2}\sqrt{r_+ r_-}\,\bigg\}^2+\frac{
B^2r^2}{4\, (1+rAx)^2}\,G(x). \label{metric}
\end{eqnarray}

Although the detailed geometry does not enter the discussion below, it is
important to identify a couple of key features. First of all, the metric
(\ref{metric})
is written in a static coordinate system, as would be natural for the
previously mentioned ``co-moving'' observers. In particular, the Killing
coordinate $s$ will be referred to as the Rindler time coordinate from
the analogy with the Rindler spacetime.

\vskip 1cm
\begin{center}
\leavevmode
\epsfysize 3in \epsfbox{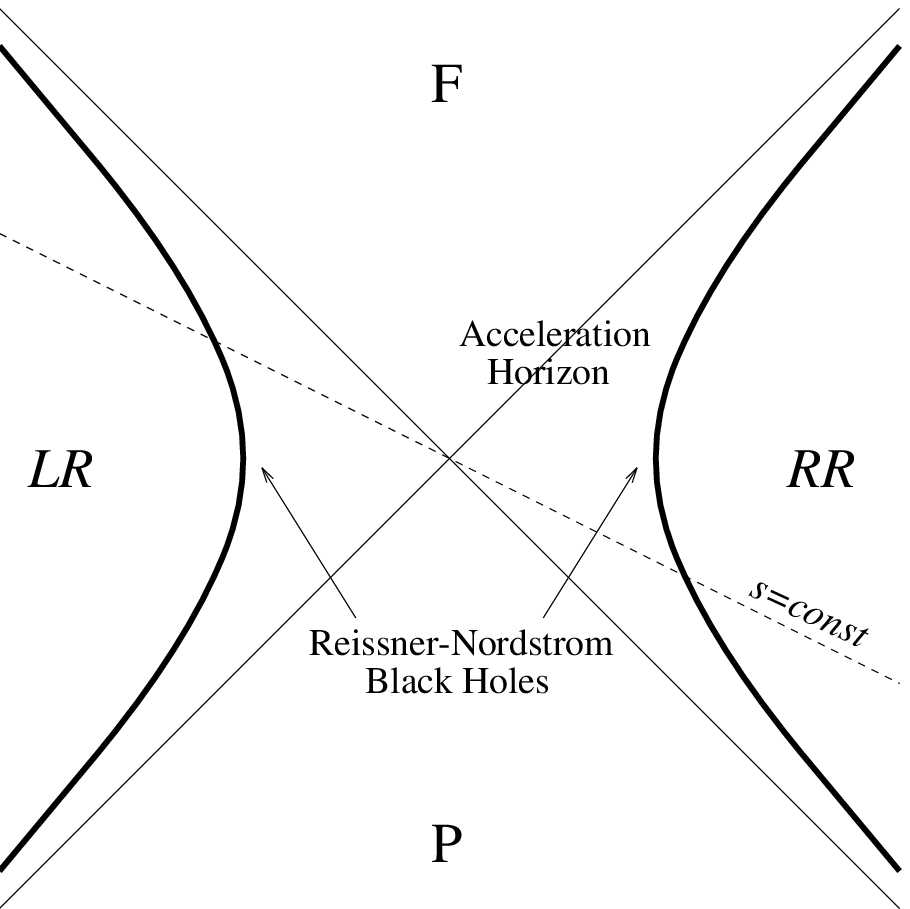}
\end{center}
\begin{quote}
{\bf Figure 1:} {\small A schematic diagram for a pair of uniformly
accelerated black holes. The black holes are represented by two
hyperbolic world lines in each Rindler wedges.}
\end{quote}

\vskip 5mm
In the limit of vanishing external electromagnetic field ($A,B\rightarrow 0$)
with fixed $r_\pm$, it is easy to see that the above metric reduces to the
more familiar Reissner-Nordstrom metric with $s$ as
the asymptotic Minkowskian time
coordinate. In this case, the geometry has two horizons at $r=r_-$ and
$r=r_+$, the latter being the black hole event horizon.

With the external
electromagnetic field that drives the acceleration, however, there
is a third, so-called acceleration horizon. Note that the same quartic
polynomial $G$ appears in all components of the metric. Call the four roots
of it, $\xi_1$, $\xi_2$, $\xi_3$, $\xi_4$ in the ascending order. Then, the
event horizon is now shifted to $r=\tilde{r}_+ \equiv-1/\xi_2A$, while
$r=r_{A}\equiv -1/\xi_3A$ is the acceleration horizon. Define the surface
gravities of the horizons:
\begin{equation}
\kappa_{BH}\equiv \frac{F'(\tilde{r}_+)}{2},\qquad
\kappa_{A}\equiv -\frac{F'(r_{A})}{2},
\end{equation}
which are related to the aforementioned temperatures by $T_{BH}=\hbar
\kappa_{BH}/2\pi$ and by $T_A=\hbar\kappa_{A}/2\pi$. See references
\cite{gauntlett}\cite{giddings}\cite{piljin} for more details on the Ernst
geometry.

As was emphasized above, I am primarily interested in the cases where the
Hawking temperature $T_{BH}$ is equal to the acceleration temperature $T_A$:
\begin{equation}
\kappa_{BH}=\kappa_{A}.\label{kappa}
\end{equation}
In some cases, most notably when the black hole mass is
much larger that its charge, $\kappa_{BH} >\kappa_A$ is always true and this
constraint can never be met. However, when the non-extremal RN black holes in
question are sufficiently close to the extremality, it is possible to achieve
this fine-tuning \cite{letter}. In fact, this constraint is naturally imposed
if the two black holes are pair-created via the wormhole-type instanton
\cite{garfinkle}.

The Euclidean version ($s\Rightarrow i\tau$)
of this metric with the  condition $\kappa_{BH}=\kappa_A$ is in fact the
instanton that induces such a tunneling event. The leading WKB exponent
from this instanton is first estimated by Garfinkle and Strominger
\cite{garfinkle}:
\begin{equation}
-\frac{S_E}{\hbar}= -\frac{\pi M^2}{\hbar |QB|}+\cdots
\end{equation}
where  $M$ and $Q$ are the mass and the charge of the pair-created black holes
respectively. The ellipsis denotes terms of higher power in $QB$.
When the size of the pair-created black holes are relatively small (
$r_\pm A \ll 1$), $M$ and $Q$ are approximately given by $M\simeq (r_++r_-)/2$
and $Q^2\simeq r_+r_-$. Also $A\simeq \kappa_A$ can be regarded as
the acceleration of the black hole and $B$ as the external
field strength that drives the acceleration. In the same limit, therefore,
the Newton's equation for the black hole motion may be written as
$MA\simeq |QB|$ $(\ll 1)$, while the fine-tuned temperature requires $r_+A
\simeq (r_+-r_-)/r_+$ $(\ll 1)$. Later on, I will come back to the leading
one-loop correction to the leading WKB exponent, again in the weak field
limit of small $QB$.

\vskip 2cm
\begin{center}
\leavevmode
\epsfysize=1.5in\epsfbox{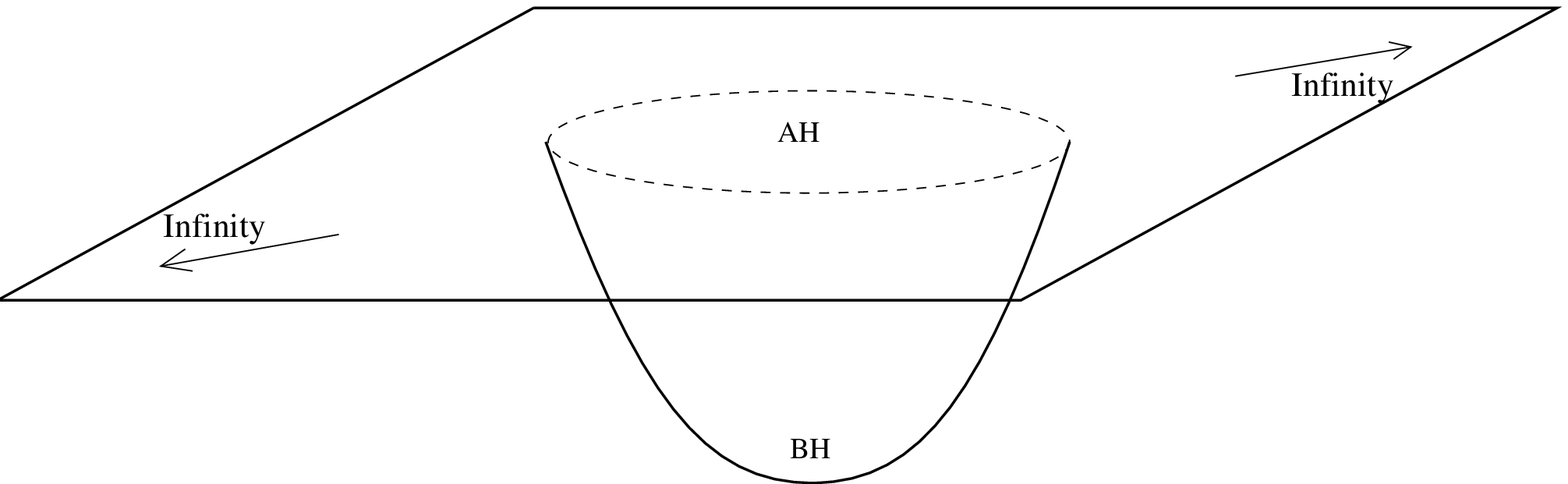}
\end{center}
\vskip 5mm
\begin{quote}
{\bf Figure 2:} {\small A schematic drawing of the pair-creation instanton.
The acceleration horizon at AH and the black hole horizon at BH are both free
of  any conical singularity thanks to the fine-tuned acceleration $\kappa_{A}
=\kappa_{BH}$. The cup-like  region is essentially a Euclidean
Reissner-Nordstrom black hole truncated beyond some large radial distance.
See ref.~\cite{piljin} for more detail.}
\end{quote}

\vskip 5mm

In addition to the above considerations, it is most essential for our purpose
that we understand the causal structure. One of the more
distinctive feature of the Ernst geometry when compared to the freely falling
black hole is the existence of the acceleration horizon, which has the
topology of $R^2$. Among other things, it complicates the asymptotic
causal structure to such an extent that the Penrose diagram cannot be drawn
on a plane. However, there is an easy way out for this extra complication:
disregard the asymptotic infinities inside Rindler wedges. The reason one
may do this is rather simple. Because the acceleration horizon
extends toward infinity along all transverse direction, a typical
future-directed quantum that originates near the black hole world line must
cross either the acceleration horizon into the asymptotic future F or
the event horizon into the black hole interior.

Thus one may visualize the causal structure relevant for the Hawking
effect by drawing a truncated Penrose diagram as  in figure 3 below.
Now it becomes clear what one must do in order to understand the difference
between the freely falling and the accelerated black holes. In quantizing
the matter field, one must take care to include the effect of the acceleration
horizon and the new asymptotic region beyond it, which can be achieved with
correct physical interpretation of various time coordinates.  Also  see
figure 5. for a comparison with a freely falling black hole.

A convenient first step in quantizing matter fields is to consider the
field equation in regions {L} and {R}. That is to say, I want to start with
the eigenmodes of the field as  appropriate for the co-moving Rindler-type
observers. In case of a freely falling black hole (depicted in figure 5.),
L and R  are simply the asymptotic regions outside the black holes,
and this is easily understood from the fact that in the absence of the
acceleration the co-moving observers at large distances are also asymptotic
inertial  observers.
\vskip 1.5cm

\begin{center}
\leavevmode
\epsfysize 4in \epsfbox{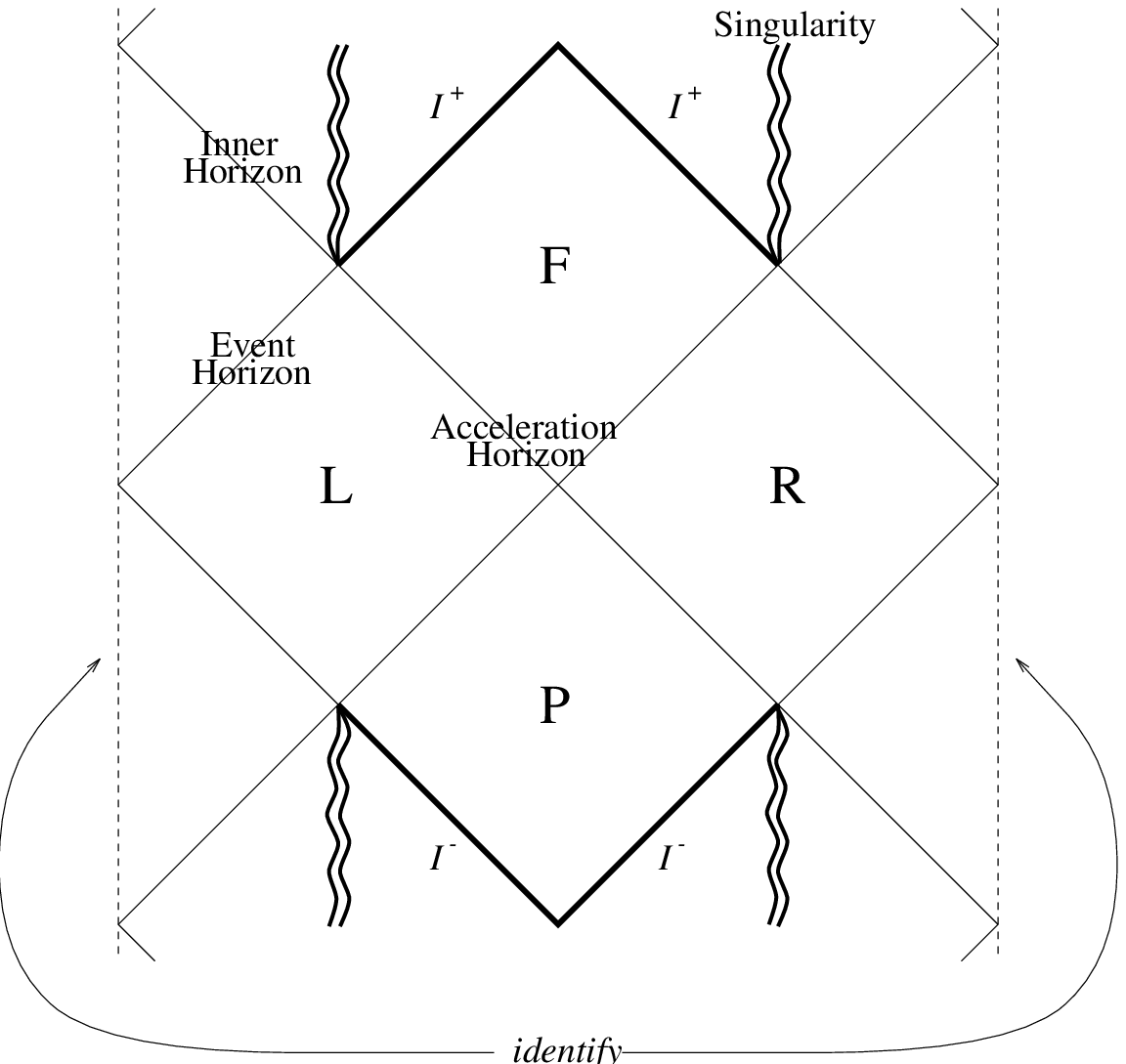}
\end{center}
\begin{quote}
{\bf Figure 3:} {\small Penrose diagram of the Ernst spacetime
with the Rindler infinities at $x=\xi_3=-1/Ar$ excised. The bold (straight)
lines indicate the asymptotic infinities.}
\end{quote}

\vskip 5mm
For this purpose, it is most convenient to introduce
a new tortoise-like coordinate $z$ between the two horizons ($\tilde{r}_+
\le r \le r_A$):
\begin{equation}
z\equiv \int^r d\tilde{r}\,\frac{1}{F(\tilde{r})},
\end{equation}
which logarithmically approaches $-\infty$ at the event horizon and $+\infty$
at the acceleration horizon.

Without loss of generality, one may consider a free scalar field with possible
quadratic curvature coupling,
\begin{equation}
\nabla^2\Psi={\cal M}^2\Psi+\cdots .
\end{equation}
After rescaling the energy and angular-momentum eigenmodes $\Psi^{(w,m)}$
for each Rindler frequency $w>0$ and the quantized angular momentum $m$,
\begin{equation}
\Psi^{(w,m)}= e^{\mp iws}\,\frac{(1+rAx)}{r}\,[\,\Phi_{(w,m)}(r,x)
e^{im\phi}\,],
\end{equation}
one finds the equation that must be solved for the eigenmodes:
\begin{equation}
w^2\Phi_{(w,m)}+\frac{\partial^2}{\partial z^2}\Phi_{(w,m)}=F(r(z))\biggl\{
\frac{1}{r^2}\biggl[-\frac{\partial}{\partial x} G(x)\frac{\partial}{\partial
x} +\frac{m^2\Lambda^4}{G(x)}\biggr]+U_{\rm eff} \biggr\}\,\Phi_{(w,m)}.
\label{eigen}
\end{equation}
Here, $U_{\rm eff}$ is a bounded function of $z$ and $x$, and in particular
contains the mass term and the possible curvature couplings.

Note that the right-hand-side of Eq.~(\ref{eigen}) has the overall factor
$F(r(z))$ that vanishes exponentially at either horizon.  Near the black
hole horizon ($z\rightarrow -\infty$), $F \sim e^{+2\kappa_{BH} z}$ while
near the acceleration horizon ($z\rightarrow \infty$), $F \sim e^{-
2\kappa_A z}$. The universal nature of the Hawking effect is related to the
fact that only those modes localized near the horizon matters, and thus
we may safely ignore such exponentially small $|z|$ dependence. Effectively,
then, a separation of variables occurs that allows one to find all the
relevant eigenmodes.

Introducing two null coordinates $u=s-z$ and $v=s+z$, one  finds the
behavior (near each horizon) of all the positive-frequency Rindler eigenmodes
$\Psi^{(w)}_L$ and $\Psi^{(w)}_R$ that have respective supports in either
{ L} or {  R}:
\begin{eqnarray}
\Psi_L^{(w)} \sim e^{-iwu}\hbox{ \ or } e^{-iwv}\quad \hbox{ in L},
& & \qquad \Psi_L^{(w)} =0 \quad \hbox{ in R }, \label{modeL}\\
\Psi_R^{(w)} \sim e^{+iwu}\hbox{ \ or } e^{+iwv}\quad \hbox{ in R},
& & \qquad \Psi_R^{(w)} =0 \quad \hbox{ in L }. \label{modeR}
\end{eqnarray}
The positive sign in (\ref{modeR}) is because $(u,v)$ grow toward past
rather than toward future in  region {R}. Suppressed here are the
dependences on the transverse coordinates $x$ and $\phi$, for these details
do not enter the discussion below.

The second step is to expand the quantum field in terms of such eigenmodes,
\begin{equation}
\Psi =\sum_{w>0} \, {\bf a}_{w}\Psi^{(w)}+{\bf a}_{w}^{\dag}\Psi^{(-w)}
\label{exp1}
\end{equation}
and declare that ${\bf a}_{w}$ and ${\bf a}_{w}^{\dag}$ span a harmonic
oscillator algebra for each eigenmode. From this one may construct the
Fock space and define the ground state or the vacuum $\vert
0\rangle$ by requiring that it be annihilated by all annihilation operators
${\bf a}_{w}$.
\begin{equation}
{\bf a}_{w}\vert0\rangle_R\equiv 0
\end{equation}
Now the field theoretical reason behind the Hawking radiation is easy
to explain. Although the above
eigenmodes are natural to certain class of observers, others
may find these modes unphysical. This ambiguity is particularly  pronounced
when there exists a horizon in the spacetime, in which case a particle state
in one mode expansion may look like an anti-particle state in another.
In the present case of the Ernst spacetime, there are at least two more
set of natural time coordinates, each of which are to be called the Kruskal
coordinates.

\vskip 1.5cm
\begin{center}
\leavevmode
\epsfysize 2in \epsfbox{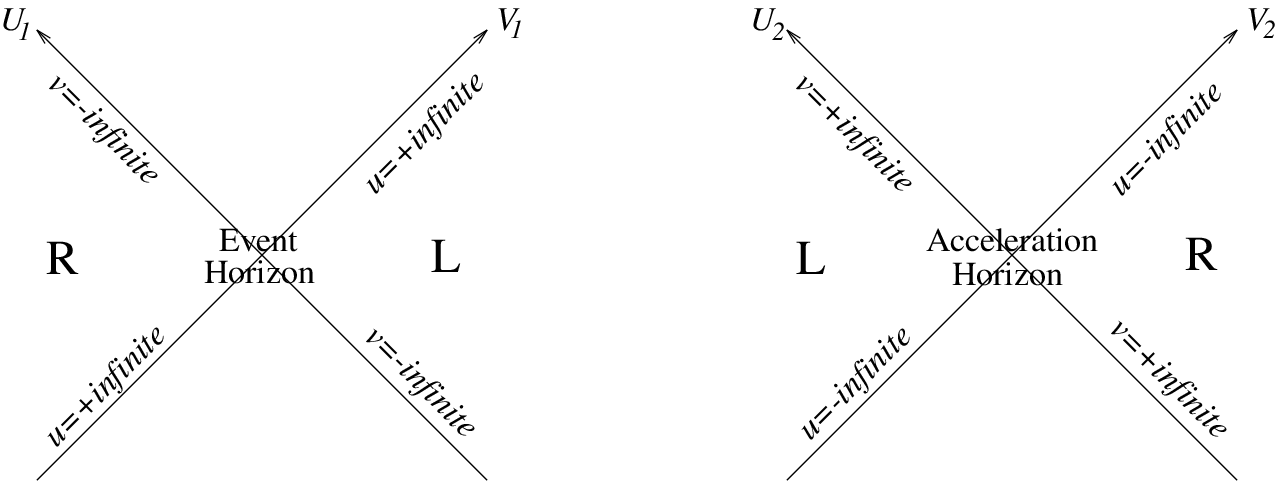}
\end{center}
\begin{quote}
{\bf Figure 4:} {\small Various null coordinates near the horizons. $U_1=0$ or
$V_1=0$ at the event horizon, while $U_2=0$ or $V_2=0$ at the acceleration
horizon. All Kruskal coordinates increase toward future. The Rindler-type
null coordinates $(u,v)$, however, increase toward future only in L,  and
actually increase  toward past in R.}
\end{quote}

\vskip 5mm

The approximate form of such coordinates
near the respective horizons, are completely determined by the surface
gravities  alone: Calling the Kruskal coordinates near the event horizon
$(U_1,V_1)$, we find,
\begin{eqnarray}
\kappa_{BH} u\simeq - \ln (-\kappa_{BH}U_1) \quad\hbox{ in L},& &\qquad
\kappa_{BH} u\simeq - \ln (+\kappa_{BH}U_1) \quad\hbox{ in R}, \label{uU}\\
\kappa_{BH} v\simeq + \ln (+\kappa_{BH}V_1) \quad\hbox{ in L},& &\qquad
\kappa_{BH} v\simeq + \ln (-\kappa_{BH}V_1) \quad\hbox{ in R}. \label{vV}
\end{eqnarray}
For the other Kruskal coordinates $(U_2,V_2)$ near the acceleration horizon,
we simply replace $\kappa_{BH}$ by $\kappa_A$ and reverse every single sign on
the right-hand-side. Already the parallel between the two horizons is manifest.

In each of these Kruskal coordinate systems, the natural mode expansion
of the quantum field $\Psi$ is distinct from the above Eq.~(\ref{exp1}): For
$(u,v)$, $(U_1,V_1)$ and $(U_2,V_2)$ respectively, one must find the
corresponding natural bases of eigenmodes,
\begin{equation}
\Psi =\sum_{w>0} \, {\bf a}_{w}\Psi^{(w)}+{\bf a}_{w}^{\dag}\Psi^{(-w)}
=\sum_{w'>0} \, {\bf b}_{w'}\Psi^{(w')}_{B}+{\bf b}_{w'}^{\dag}
\Psi^{(-w')}_{B} =\sum_{w''>0} \, {\bf c}_{w''}\Psi^{(w'')}_A+{
\bf c}_{w''}^{\dag}\Psi^{(-w'')}_A ,
\end{equation}
which lead to the natural vacuum appropriate for each coordinate system:
\begin{equation}
{\bf a}_{w}\,\vert  0\rangle_R\equiv 0,\qquad
{\bf b}_{w'}\,\vert  0\rangle_{B}\equiv 0,\qquad
{\bf c}_{w''}\,\vert  0\rangle_A\equiv 0.
\end{equation}
The ``black hole vacuum'' $\vert  0\rangle_{B}$ is the one that allows smooth
event horizon(s), while the ``asymptotic vacuum'' $\vert  0\rangle_A$ allows
smooth acceleration horizon(s). Depending on the precise initial conditions
to be imposed on the quantum field, the physical vacuum would be given by
either the former only or a certain composite of the two. The static vacuum
composed of $\vert 0\rangle_{B}$ type only is the physical vacuum for the
Euclidean instanton geometry and in the limit of $\kappa_A=0$
corresponds to the so-called Hartle-Hawking vacuum.

In general, different mode expansions are related by unitary
transformations. When the unitary transformation in question  do not mix
creation operators with annihilation operators, the transformation would act
trivially on the vacuum state itself. However, in the present case with
horizons, the above three vacua are expected to be
inequivalent, for the relevant unitary transformation, often called Bogolubov
transformations,  mixes in negative and positive modes rather
indiscriminately. For instance, following Unruh  \cite{unruh} one
may choose the following unitary transformation rule to construct
eigenmodes $\Psi_B$'s that are appropriate near the black hole event horizon,
\begin{equation}
{\Psi}^{(w)}_{BL}\simeq N_w(\kappa_{BH})[\Psi^{(w)}_L+e^{-\pi w/\kappa_{BH}}
\Psi^{(-w)}_R],
\qquad
{\Psi}^{(w)}_{BR}\simeq N_w(\kappa_{BH})[\Psi^{(w)}_R+e^{-\pi w/\kappa_{BH}}
\Psi^{(-w)}_L],
\label{RB}
\end{equation}
with $N_w(\kappa_{BH})\equiv 1/\sqrt{1-e^{-2\pi w/\kappa_{BH}}}$. This
leads to a specific relationship between the two vacua:
\begin{equation}
\vert 0\rangle_B={\cal S}(\kappa_{BH})\,\vert 0\rangle_R . \label{bhv}
\end{equation}
The  Bogolubov transformation ${\cal S}(\kappa_{BH})$ depends on the single
parameter $\kappa_{BH}$ or equivalently $T_{BH}$, and excites the Rindler
modes $\Psi_L$ and $\Psi_R$ in a pairwise and thermal fashion.

\vskip 1.5cm

\begin{center}
\leavevmode
\epsfysize 4in \epsfbox{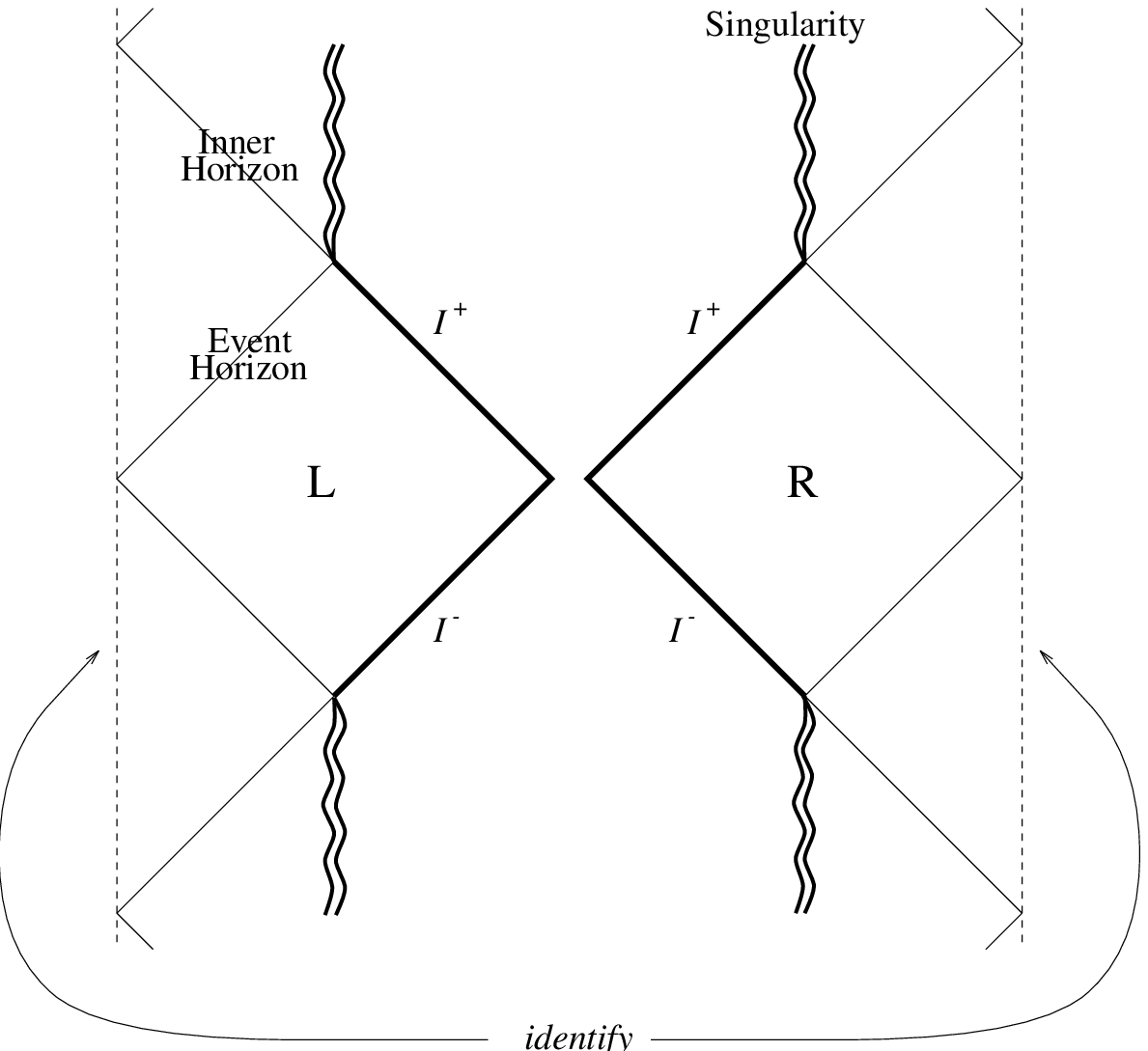}
\end{center}
\begin{quote}
{\bf Figure 5:} {\small Penrose diagram of the ``freely falling''
Reissner-Nordstrom black hole with a positive Hawking temperature
($\kappa_A=0$, $\kappa_{BH}\neq0$).
The Hawking effect induces a thermal radiation toward the asymptotic
future infinities $I^+$.}
\end{quote}

\vskip 5mm
For an ordinary nonaccelerated black holes ($\kappa_A=0$),
this would be the end of the story, since $(u,v)$ are
themselves the asymptotic Minkowskian coordinates. In this limit of
$\kappa_A=0$, one finds $\Lambda\equiv 1$ and $F\rightarrow 1$ at large
distances, so that the metric is approximately given by $g\simeq -dudv+\cdots$.
The second Kruskal coordinate system $(U_2,V_2)$ is irrelevant since the
acceleration horizon does not exist, as illustrated in figure 5. For physical
black holes with smooth future event horizon, then, the physical vacuum
known as the Unruh vacuum is such that the asymptotic inertial observers
find outward thermal radiation at $T_{BH}=\hbar\kappa_{BH}/2\pi$
\cite{hawking}.

However, with the uniformly accelerated black holes, $(u,v)$ are not
asymptotic inertial coordinates. Rather, $(U_2,V_2)$ are, or more precisely
certain coordinates that behaves as $(U_2,V_2)$ near the acceleration horizon.
It is particularly easy to see this in the limit of small $QB$ when the black
hole size is relatively small compared to the Schwinger length. Introducing
a new set of coordinates far away from the black hole:
\begin{equation}
A^2\zeta^2\simeq \frac{1-r^2A^2}{(1+rAx)^2},\qquad
A^2\rho^2\simeq A^2r^2\frac{1-x^2}{(1+rAx)^2},
\end{equation}
it leads to the following approximate form of the Ernst metric far away
from the black holes,
\begin{eqnarray}
g &\simeq& \Lambda^2\,( -A^2\zeta^2\,ds^2+d\zeta^2)+\Lambda^2\,d\rho^2
+\Lambda^{-2}\,
\rho^2\,d\phi^2,\qquad \Lambda\simeq 1+\frac{B^2\rho^2}{4}\nonumber \\
&\simeq & \Lambda^2\,(-dU\,dV)+\cdots. \nonumber
\end{eqnarray}
Here the null Minkowskian coordinates $(U,V)$ are defined as $U= +\zeta
e^{+As}$ and $V= -\zeta e^{-As}$ in region L and to be analytically continued
everywhere else outside the black  hole.
Recalling that $A\simeq  \kappa_A$ in this weak field
limit and that $2\kappa_A z\simeq -\ln F \simeq -2\ln A \zeta$ as $\zeta
\rightarrow 0$, one finds that the null
Minkowskian coordinate system $(U,V)$ coincides with $(U_2,V_2)$ near
the acceleration horizon. Since $\Lambda^2$ is never zero, this observation
implies among other thing that $\vert 0\rangle_A$ is not only a natural
vacuum near  the acceleration horizon but the asymptotically empty,
Minkowskian vacuum. More  general form of the coordinates $\zeta$ and $\rho$
suitable for all values of $QB$, can be found in Ref.~\cite{giddings}.

Now one must perform another Bogolubov transformation near the acceleration
horizon at $\zeta\simeq 0$.
In fact the situation is exactly  parallel to the above, except
that the relative positions of { L} and  {R} are switched. For the asymptotic
inertial modes $\Psi_A^{(w)} $'s, we find near the acceleration horizon:
\begin{equation}
{\Psi}^{(w)}_{AR}\simeq N_w(\kappa_A)[\Psi^{(w)}_R+e^{-\pi w/\kappa_A}
\Psi^{(-w)}_L],
\qquad
{\Psi}^{(w)}_{AL}\simeq N_w(\kappa_A)[\Psi^{(w)}_L+e^{-\pi w/\kappa_A}
\Psi^{(-w)}_R].
\label{RA}
\end{equation}
This again leads to the following relationship between the two vacua:
\begin{equation}
\vert 0\rangle_A={\cal S}(\kappa_{A})\,\vert 0\rangle_R . \label{asv}
\end{equation}
A comparison with (\ref{bhv}) makes immediate the
special nature of fine-tuned geometry  $\kappa_A=\kappa_{BH}$. As seen by
the co-moving Rindler observers (to whom $\vert 0\rangle_R$ is the natural
empty vacuum), both the black hole vacuum $\vert 0\rangle_B$ and the
asymptotic $\vert 0\rangle_A$ appear thermal, which results in an
equilibrium when  the two temperatures happen to be equal.
On the other hand, for the asymptotic inertial observers (to whom $\vert 0
\rangle_A$ is the natural empty vacuum), the natural black hole vacuum
may be written as
\begin{equation}
\vert 0\rangle_B  ={\cal S}(\kappa_{BH})\,\vert 0\rangle_R
={\cal S}(\kappa_{BH})\,{\cal S}(\kappa_{A})^{-1}\vert 0\rangle_A
\quad \rightarrow \quad \vert 0\rangle_A\qquad\hbox{if $\kappa_A=\kappa_{BH}$}.
\label{main}
\end{equation}
Therefore, unlike in the case of a freely falling black hole, the
black hole vacuum $\vert 0\rangle_B$ here is in fact equivalent to the
asymptotically empty vacuum $\vert 0\rangle_A$ and there is no possibility of
a Hawking-type radiation toward the asymptotic inertial observers.

\vskip 5mm
A couple of remarks are in order. First of all, I have used the eternal
Ernst geometry where both future  horizons and past horizons are taken
seriously, while in more realistic geometries the past horizons would be
absent. However, we believe that this idealization does not alter the final
conclusion, in much the same way one may derive the Bogolubov transformation
responsible for the Hawking radiation using eternal black holes  rather than
realistic ones without the past event horizon. Note that the matter of the
initial condition is obviated in the present problem because of the null
Bogolubov transformation.

This can be seen more clearly by considering the transformation
between  the two Kruskal coordinates:
\begin{equation}
U_2\sim -\frac{1}{U_1}, \qquad V_2\sim -\frac{1}{V_1}.\label{ex}
\end{equation}
This transformation is easily seen to preserve the lower-half-planes of the
Kruskal time coordinates, being an $SL(2,R)$ generator, which in turn
explains why the Bogolubov transformation thereof is trivial to the leading
approximation. In fact, this transformation (\ref{ex}) is exactly what we
would have found if we had been considering a freely falling extremal RN
black hole that has zero Hawking temperature and thus no late-time Hawking
radiation, provided that we again identify $U_2$ and $V_2$ as the asymptotic
inertial time coordinates.

For the extremal black hole, however, the above argument that
utilizes the property of $SL(2,R)$ transformations is not entirely correct
since the relevant range of $U_1$ for instance should be confined to the
negative real line. Nevertheless, the
fact that the extremal black hole is of zero Hawking temperature remains true:
While the resulting Bogolubov transformation is not entirely trivial, it
does not involve a steady flow of radiation energy either. In a similar
vein, we expect the present coordinate transformation in (\ref{ex})
lead to the vanishing Hawking radiation for the single accelerated black hole
formed by gravitational collapse, although only half-lines of the Kruskal
coordinates would overlap with each other in that case.

Also, this along with the fact the above transformation laws are valid only
near the respective horizons, tells us that there could be certain transient
behavior: More careful analysis could predict some residual one-loop effect
even when the leading Hawking flux vanishes. For instance, in the absence of
unbroken extended local supersymmetry, the extremal Reissner-Nordstrom
black hole of vanishing Hawking temperature may suffer a finite energy loss,
which shifts its mass by a small amount  $\sim \hbar/Q$ \cite{jaemo}.
Similarly one should expect a  similar mass shift for the
present non-extremal black holes under the uniform acceleration, which
actually manifests itself in the one-loop corrected
tunneling rate of the black hole pair-creation as will be shown shortly.

\vskip 5mm
Finally  we are in position to discuss the pair-creation problem. As was
briefly mentioned above, the Euclidean version of the above geometry is the
instanton that mediates this quantum tunneling. While general one-loop
WKB tunneling rate would be rather difficult to obtain  owing to the
uncertainty in the gravitational sector, it is in principle possible to
calculate the one-loop  contribution from the matter fluctuations.
Then, the problem reduces to estimating certain matter
partition functions in the background of Euclidean Ernst metric with $T_{BH}
=T_A$.

In principle one would try to perform a (Euclidean) mode expansion of the
relevant functional determinant, but this approach is unlikely to be effective.
Even for the far simpler case of freely falling Reissner-Nordstrom
black hole geometry, such a program was carried out only very recently
\cite{hiscock}.
Alternatively one may concentrate on the weak field behavior of
one-loop correction and consider just the
leading $QB$-dependence of the additive one-loop correction $W$ to the leading
WKB exponent $-S_E/\hbar$. Varying with respect to the background geometry and
using the definition for the energy-momentum expectation values, one gets
\begin{equation}
-\delta W= \int dx^4\,\sqrt{g}\;\delta g^{\alpha\beta}\,\langle 0\vert
T_{\alpha\beta}\vert 0\rangle_{\rm one-loop}.
\end{equation}
It is most convenient to keep the external field $B$ fixed (or equivalently the
temperature $T_{BH}=T_A$) and vary with respect to the black hole charge $Q$.
A crucial point here is that the instanton geometry has only two independent
(dimensionful) parameters.

This integral picks up a trivial factor of $\hbar/T_{BH}\simeq 1/B$ from the
periodicity of the Euclidean time spanned by the Killing coordinate
$\tau=-is$, and therefore the remaining spatial part of the integral
determines the leading $QB$-dependence of $W$. Thus the behaviors of one-loop
energy-momentum expectation values in various spatial regions become
a matter of essential importance.

First of all, since the local Euclidean geometries
are given by $S^2\times D^2$ near the black hole horizon and by $R^2\times
D^2$ near the acceleration horizon, one should not expect any singular
behavior of $\langle 0\vert T_{\alpha\beta}\vert 0\rangle_{\rm one-loop}$
there.\footnote{As it turned out, there is some subtlety in going to the
$QB\rightarrow 0$ limit, which was shown to be harmless at least for
conformally coupled fluctuations \cite{piljin}.}

What about the asymptotic region? While one do not expect singular
$\langle 0\vert T_{\alpha\beta}\vert 0\rangle_{\rm one-loop}$
there either, even such a mild behavior as $\langle 0\vert T_{\alpha\beta}
\vert 0\rangle_{\rm one-loop}\rightarrow $ constant, is dangerous due to the
infinite volume associated with the region.
In fact, if we were considering a ``freely falling'' Euclidean black
hole geometry, the vacuum with smooth black hole horizon would be
intrinsically thermal at large spatial distances: The asymptotically
constant energy-momentum expectation value thereof would induce a huge
gravitational backreaction that distorts the geometry at large distances.
And this is where the main result (\ref{main}) makes the difference.

Unlike the case of a ``freely falling'' Euclidean black hole, the instanton
geometry is such that the natural Hartle-Hawking type vacuum with smooth
horizons is asymptotically trivial: $\langle 0\vert T_{\alpha\beta}\vert
0\rangle_{\rm one-loop}$ vanishes rapidly far away from the Euclidean black
hole horizon. It is only inside the truncated Euclidean black hole region
(see figure 2.) that the vacuum state appears thermal. At the moment it is
unclear how rapidly the energy-momentum vanishes at large distances.
But the point is, the possible gravitational backreaction to the
one-loop quantum effect is a far less serious problem than one might have
anticipated otherwise.

Now a reasonable conjecture would be that the behavior of the one-loop
energy-momentum is asymptotically insensitive to the truncated Euclidean
black hole at the center and thus is asymptotically identical to that of
the one-loop energy-momentum in the background Melvin space without any
black hole. Then it suffices to consider the above integral over the
truncated Euclidean black hole, as argued in Ref.~\cite{piljin},  for it
is the difference between two partition functions on the instanton and the
background geometry that enters the pair-creation rate. In the weak
field limit $QB\rightarrow 0$ where the black hole mass $M$ is equal to
the charge  $|Q|$ to the leading order in $QB$, a simple dimensional
argument can be used to show that
\begin{equation}
W\sim\frac{1}{QB}\quad\Rightarrow\quad -\frac{S_E}{\hbar}+W\simeq
-\frac{\pi M^2}{\hbar |QB|}\left\{1+\sigma\frac{\hbar}{Q^2}\right\}+\cdots
\end{equation}
The proportionality constant $\sigma$ was explicitly calculated for the
chargeless sector of Callan-Rubakov modes, each of which contributes
$-1/36\pi$. See Ref.~\cite{piljin} for more detail.

The only possible interpretation of this one-loop correction seems to be that
the effective mass of the pair-created black hole is shifted from $M\simeq |Q|
$ to $M_{semi}\simeq |Q|\,\{1+\sigma\hbar/2Q^2\}$. It is as if the external
magnetic field $B$ creates a pair of particles with the charges $\pm Q$
but the mass $M_{semi}$ instead of $M$. How does this fit into the previously
known one-loop effects on Reissner-Nordstrom black holes?

We have found that the pair-created black holes do not suffer from the usual
Hawking radiation as long as the fine-tuned acceleration is maintained.
However, as briefly mentioned above, this does not mean that the one-loop
effect is completely absent, rather this implies that the one-loop effect
may induce at most a finite shift of the black hole mass.

A similar circumstance exists for a freely falling extremal Reissner-Nordstrom
black hole. The leading late-time Hawking radiation vanishes due to the
vanishing Hawking temperature but there
are in general subleading transient radiation of finite integrated flux.
Again the mass shift was explicitly calculated \cite{jaemo} for chargeless
sector of the Callan-Rubakov modes:
\begin{equation}
\frac{\Delta M}{M}=-\frac{N\hbar}{72\pi Q^2},
\end{equation}
where $N$ is number of the chargeless Callan-Rubakov modes. This result
is easily seen to be consistent with the above value of $\Delta\sigma
=-1/36\pi$ from each chargeless Callan-Rubakov modes in the black
hole pair-creation rate.

(It is worthwhile to recall  that the sum total of such one-loop effects
and thus the total $\sigma$ are expected to vanish identically if the theory
is embedded in certain unbroken extended supergravity, due to the Bogomol'nyi
bound interpretation of the extremal black hole \cite{gibbons}.)

The point I want to emphasize here is not so much that these corrections
are found, as that the naive WKB procedure for the tunneling rate
seems to work pretty well. Not only the potential problem from
the gravitational backreaction turned out to be rather benign, but the
resulting one-loop corrections above are quite consistent with another
one-loop effect, which was already found and estimated rigorously in
Ref.~\cite{jaemo} and which involves the far simpler background geometry of
freely falling extremal black hole. This renders more weight to the validity
of the semiclassical method in the pair-creation process.

\vskip 5mm

Before closing, it is appropriate to return to the classic puzzle of the
Bremmstrahlung from a uniformly accelerated charge. Clearly the nature of
the radiations in the two problems are too different to allow any naive
comparison to be made as above. But at the same time, it is instructive
to understand exactly where the key differences lie.

The energetics part of the puzzle was in fact first understood by Coleman
\cite{coleman}
almost twenty years before Boulware's conclusive work. The crucial observation
by Coleman was that one must take care to include the energy associated with
the (boosted) Coulomb field around the moving charge. After an appropriate
regularization of the point-like charged particle, the total energy of the
system may be split into three pieces: the kinetic energy of the charged
particle, the radiation energy of the Bremmstrahlung, the electromagnetic
energy of the Coulomb field. In effect, the last acts as a sort of energy
reservoir that mediates the energy transfer from the first to the second and
{\it in the special case of uniform acceleration provides all the radiation
energy without extracting any from the charged particle. }

Furthermore, whenever the
acceleration lasts only for a finite duration, the initial  and the final
Coulomb fields are identical up to a boost and the energy conservation between
the charged particle and the Bremmstrahlung is well maintained:
\begin{equation}
\Delta E=\int^f_i dt\, \frac{2e^2}{3}{\bf A}^2+\int^f_i dt\,{\bf v \cdot
F}_{\rm damping}= \frac{2e^2}{3}\int^f_i dt\,\frac{d}{dt}{\bf v \cdot A}
=\frac{2e^2}{3}{\bf v \cdot A}\biggr\vert^f_i\quad\rightarrow \quad 0.
\end{equation}
Thus the existence of the Bremmstrahlung in the inertial frame is perfectly
consistent with the energy conservation. I refer the reader to
Ref.~\cite{boulware} for the complete resolution of this classic problem.

In comparison, a nonvanishing Hawking radiation with $T_{BH}=T_{A}$ would
have been very difficult to explain despite the (superficial)
similarity from the viewpoint of co-moving
observers. Simply put, there is no such intermediary as the Coulomb field
that could explain the different energy flows that should have been seen by
different classes of observers. There was always a logical possibility that
the subtlety in defining the black hole mass in the asymptotically nonflat
geometry of the Ernst metric might play a role, but it is gratifying to know
that the simplest possible answer is also true.

\vskip 1cm
\centerline{\bf Acknowledgement}
I am grateful to the organizers of the Sixth Moscow Quantum Gravity Seminar
for their hospitality.  This work is supported in part by the U.S. Department
of Energy.


\vskip 1cm
\begin{thebibliography}{99}
\bibitem{hawking}{S.W. Hawking, Comm. Math. Phys. 43 (1975) 199.}
\bibitem{bath}{S.A. Fulling, Phys. Rev. D7 (1973) 2850; P.C.W. Davies,
J. Phys. A8 (1975) 609.}
\bibitem{unruh}{W.G. Unruh, Phys. Rev. D14 (1976) 870.}
\bibitem{letter}{P. Yi, Phys. Rev. Lett. 75 (1995) 382;
{\it Quantum Stability of Accelerated Black Holes},
CU-TP-690,  hep-th/9505021.}
\bibitem{giddings}{F. Dowker, J.P. Gauntlett, S.B. Giddings,  G.T.
Horowitz, Phys. Rev. D50 (1994) 2662.}
\bibitem{boulware}{D.G. Boulware, Ann. Phys. (New York) 124 (1980) 169:
H. Ren and E. Weinberg, Phys. Rev. D49 (1994) 6526.}
\bibitem{quantum}{
A. Higuchi, G.E.A. Matsas and D. Sudarsky, Phys. Rev. D45 (1992) R3308.}
\bibitem{garfinkle}{D. Garfinkle and A. Strominger, Phys. Lett. 256B (1991)
146.}
\bibitem{piljin}{P. Yi, Phys. Rev. D51 (1995) 2813.}
\bibitem{ernst}{F.J. Ernst, J. Math. Phys. 17 (1976) 515.}
\bibitem{gauntlett}{F. Dowker, J. P. Gauntlett, D. A. Kastor and J. Traschen,
Phys. Rev. D49 (1994) 2909.}
\bibitem{hiscock}{P.A. Anderson, W.A. Hiscock and D.A. Samuel,
Phys. Rev. Lett. 70 (1993) 1739; references therein.}
\bibitem{jaemo}{J. Park and P. Yi, Phys. Lett. 317B (1993) 41.}
\bibitem{gibbons}{G.W. Gibbons and C.M. Hull, Phys. Lett. 109B (1982) 190.}
\bibitem{coleman}{S. Coleman, Rand Report RM-2820 (1961), unpublished.}
\end{thebibliography}
\end{document}